# Interplanetary transmissions of life in an evolutionary context

Ian von Hegner
Aarhus University

**Abstract** The theory of lithopanspermia proposes the natural exchange of organisms between solar system bodies through meteorites. The focus of this theory comprises three distinct stages: planetary ejection, interplanetary transit and planetary entry. However, it is debatable whether organisms transported within the ejecta can survive all three stages. If the conjecture is granted, that life can indeed be safely transmitted from one world to another, then it is not only a topic pertaining to planetary science but also biological sciences. Hence, these stages are only the first three factors of the equation. The other factors for successful lithopanspermia are the quality, quantity and evolutionary strategy of the transmitted organisms. When expanding into new environments, invading organisms often do not survive in the first attempt and usually require several attempts through propagule pressure to obtain a foothold. There is a crucial difference between this terrestrial situation and the one brought about by lithopanspermia. While invasive species on Earth repeatedly enters a new habitat, a species pragmatically arrives on another solar system body only once; thus, an all-or-nothing response will be in effect. The species must survive in the first attempt, which limits the probability of survival. In addition, evolution sets a boundary through the existence of an inverse proportionality between the exchanges of life between two worlds, thus further restricting the probability of survival. However, terrestrial populations often encounter unpredictable and variable environmental conditions, which in turn necessitates an evolutionary response. Thus, one evolutionary mode in particular, bet hedging, is the evolutionary strategy that best smooth out this inverse proportionality. This is achieved by generating diversity even among a colony of genetically identical organisms. This variability in individual risk-taking increases the probability of survival and allows organisms to colonise more diverse environments. The present analysis to understand conditions relevant to a bacterial colony arriving in a new planetary environment provides a bridge between the theory of bet hedging, invasive range expansion and planetary science.

**1. Introduction**

E. de Montlivault (1821) suggested that meteorites could be carriers of life to Earth, while Hermann Richter (1865) proposed a hypothesis detailing that, rather than falling to the ground as meteorites, some meteors instead can glance the Earth's atmosphere at such an angle that they penetrate only partially and potentially collect airborne organisms along the way, before bouncing back into space with them. Since then, lithopanspermia (rock panspermia), the theory that proposes the natural exchange of organisms between solar system bodies due to asteroidal or cometary impactors [Nicholson, 2009], has become an active field of research.

Scientific research involving lithopanspermia has focused on the following three distinct stages: (1) Planetary ejection – organisms need to survive ejection from a planet; (2) Interplanetary transit – organisms need to survive the transmission through space; and (3) Planetary entry – organisms needs to survive entry from space onto a planet. These stages encompass multidisciplinary research, ranging from studies of ejection rates to dynamical transfer and impact physics, which seeks to clarify whether life can survive each of these stages and estimate the overall chances of survival when organisms are transferred between different solar system bodies.

One of the early objections to lithopanspermia was based on a doubt as to whether material could even be exchanged between solar system bodies. However, since its proposal, it has become evident that among the over 62,000 meteorites found on Earth [The Meteoritical Society, 2020], 142 are Martian in origin [Irwing, 2020], eliminating any doubts regarding the exchange of material between planets. Most Martian meteorites have been on Earth long before they were found, although some were found after contact, such as the Tissint meteorite that fell to Earth near Morocco on 18 July 2011 [Brennecka et al., 2014]. The opposite situation, in which terrestrial material arrives on Mars, seems conceivable.

It is estimated that the escape velocity, which is the minimum velocity required to escape a gravitational field, is 5.03 km/sec for Mars and 11.2 km/sec for Earth [Nicholson, 2009]. Thus, the stages of lithopanspermia would work more easily on planets like Mars. However, Beech et al. (2018) estimated that at least $10^{13}$ kg of material has been ejected from Earth into the inner solar system since the beginning of the Phanerozoic Eon, with an estimated 67% attributed to the Chicxulub cratering event. This terrestrial material could potentially reach Mars as well as Venus provided an impact velocity of 25 km/s, and could potentially reach the moons of Jupiter as well, provided the impact velocity exceeds 27 km/s. This estimate was based on documented terrestrial impact craters with diameters larger than 5 km. It is clear that undocumented craters could increase the estimate of the matter ejected from earth to perhaps a magnitude on the order of ~$10^{14}$ kg, which the researchers deduced on the basis of the documented craters.

However, these travel times are, from a biological point of view, often long. The majority of Martian meteorites are estimated to have spent from approximately ~0.6 to 15 million years in space. For example, the Martian meteorite Nakhla that fell on 28 June 1911 in northern Egypt had a transmission time in space of approximately ~10.75 million years [Nyquist et al., 2001], and the Tissint meteorite is estimated to have been ejected between 700,000 and 574 million years ago [Brennecka et al., 2014]. However, based on calculations, shorter travel times may be possible, with some meteorites potentially reaching transit trajectories fast enough to allow interplanetary travel lasting months to a few years between Mars and Earth [Melosh, 1988], although it should be noted that major shocks at launch and impact would reduce the chance of organism survival.

Since lithopanspermia conjectures that ejected rocks serve as transporters of organisms, and considering that most Martian meteorites are igneous, it is imperative to investigate the presence and persistence of organisms in igneous terrestrial rocks, such as basalts and granites. In one study, microbial communities were recovered from eleven rock samples collected from the walls of deep subsurface tunnels at Rainier Mesa of the Nevada Test Site in the United States. The total numbers of microorganisms per gram of rock from these samples were approximately $4 \times 10^5$ to $4 \times 10^7$ [Haldeman et al., 1994].

Benardini et al. (2003) reported the isolation of culturable endolithic *Bacillus* sp. from near-subsurface basalt rocks collected in the Sonoran desert near Tucson, Arizona (USA); however, the number of spores were exceedingly low (approximately ~10 spores per gram of rock). In contrast, a related study obtained markedly higher numbers of cultivable endolithic *Bacillus* spores ($5 \times 10^2$ spores from approximately out of ~$10^4$ cultivable bacteria per gram) from the interior of near-subsurface granite rocks, also collected in the Sonoran desert. The authors noted that these isolates were closely related to a few *Bacillus* species that inhabit globally distributed endolithic sites and some extreme environments [Fajardo-Cavazos and Nicholson, 2006].

Thus, life does exist in globally distributed rocks that can potentially be ejected. Nevertheless, the probability of their survival in the stages of lithopanspermia remains in question. Many studies have addressed this issue, which involves not only surviving the launch and landing, but also surviving the transmission. During the free exposure to space during transport, the organisms are subjected to intense radiation, and experience complete desiccation, rendering them unable to maintain metabolic functions and perform cellular repair, which leads to degradation of the biological structure and death of the organisms [Mileikowsky, et al., 2000].

However, bacterial endospores have long been studied with regard to space conditions. These spores are the toughest terrestrial life form known, and are capable of enduring extreme conditions in the dormant state that otherwise damage or are lethal to other known forms of life (Nicholson et al., 2000). Calculations by Mileikowsky et al. (2000) predicted that a fraction of a *Bacillus subtilis* spore population ($10^{-6}$) would in fact survive cosmic radiation in space for approximately 1 Ma if protected by 1 m of meteorite material (assuming the rock initially harbours about $10^8$ spores/g) and 25 Ma if protected by 2 to 3 m of meteorite material. Thus, the natural transfer of microorganisms such as *Deinococcus radiodurans* and *Bacillus* sp. was concluded as not only possible but highly probable in Mars–Earth interplanetary travel. Interestingly, a study of many Martian meteorites indicated that some never experienced heating temperatures higher than

approximately100 °C during their ejection from Mars and transfer to Earth [Shuster and Weiss 2005], which was one of the criteria of survivability set by Mileikowsky et al. (2000).

It is important to mention that the key mechanism behind lithopanspermia is not restricted to the past, but is still active in the solar system [Beech et al., 2018]. An example is the aforementioned 2011 arrival of the Tissint meteorite [Brennecka et al., 2014], demonstrating that meteorites continue to be transmitted between solar system bodies to the present-day.

Despite of all this, lithopanspermia may still be considered an extreme situation from a biological point of view. Whether the organisms transported within the ejecta can survive all the three aforementioned stages is being actively investigated. Although only a few different organisms have been studied, new research data continue to be generated. The current consensus appears to be that organisms may indeed survive all three stages of lithopanspermia [Beech et al., 2018].

The notions that life can survive transmission to another world and that life has actually been transmitted to another world are of course two different situations. We do not yet know if life or biotic material has ever been transmitted between worlds in this solar system. Although the probability of a successful transfer of life between worlds seems modest and the probability of establishing a viable ecosystem in a new world seems even more modest, the possibility nevertheless seems to exist. So if the conjecture is granted that life can indeed be transmitted safely from, for example, Earth to Mars or vice versa, then it is not only primarily a topic of planetary science but also a biological topic.

Hence, planetary ejection, interplanetary transit and planetary entry are only the first three factors in the equation. After all, the defining characteristic of life is not chemistry or physics, but evolution. Thus, it is important to clarify the evolutionary context of lithopanspermia. Thus, organisms are adapted to their environment. An organism that does well in one environment will not do so well in another. Thus, organisms cannot simply be transmitted to another environment since they will encounter severe environmental stressors there.

Although many studies have investigated the possible process by which life may be transmitted and the conditions required for their survival through the journey, relevant systematic and predictive evolutionary theories are still lacking. In this article, I seek to clarify how the arrival of life on another world can be understood in an evolutionary framework and what strategy organisms must follow to survive the encounter with that world.

## 2. Discussion

The survival of organisms arriving at another solar system body is more complex than just their survival during the actual transport. Planetary ejection, interplanetary transport and planetary entry are just the first three factors in the equation for successful lithopanspermia. The other factors are the quality, quantity, and evolutionary strategy of the transported organisms.

The focus here will be on the last factor, the evolutionary strategy. This concerns the situation that applies if life successfully reaches another world and its interpretation in an evolutionary framework. The many variables involved in lithopanspermia make it uncertain what probability life has in surviving and gaining a foothold on another solar system body. However, uncertainty is in fact a prevalent characteristic of terrestrial environments, which in turn necessitates evolutionary responses. I will make the following simplifying assumptions in my discussion:

(i) It is assumed that the meteorite manages to bring living organisms to another world.
(ii) It is assumed that the meteorite lands where the transported organisms have a possibility of survival.
(iii) It is assumed that the meteorite carries with it members of the same clonal colony.

Given the general toughness and versatility of bacteria compared to other terrestrial species, and the fact that they are the dominant species on this world, existing in all available niches, the likelihood that bacteria will be picked up and survive the journey is greatest. Hence, I will focus on bacteria. In sections 2.1, 2.2 and 2.3, I discuss the 3 arrival modes that lithopanspermia can be divided into. These are fractional genetic transfer, fractional componental drift and phenotypic continuance. In subsection 2.3.1, I discuss the evolutionary framework for the survival of organisms on another world that emerges through the 3 overarching evolutionary modes of response to changing environments.

**2.1. Fractional genetic transfer**

Research into the possibility of lithopanspermia has focused on 3 distinct stages that bring living organisms to another solar system body. However, perhaps initially surprising, such entrapped organisms do not necessarily have to survive the journey in order to have an astrobiological significance. The meteorite need not be able to keep them alive. In fact, a meteorite only needs to bring its content, including the organism, to the acceptor world. This is because even if the organisms do not survive the journey, they can still potentially influence any biological evolution that takes place in the environment in which they arrive.

Competent bacterial species can actively take up exogenous DNA, sometimes termed naked DNA fragments, from the surrounding environment and occasionally incorporate it into their chromosomes through homologous recombination, thus changing their own genotype [Mell et al., 2011]. This well-known mechanism, termed transformation, is the primary mode of horizontal gene transfer in bacteria, along with conjugation and transduction [Blokesch, 2016]. Thus, it is in principle possible for a hypothetical Martian bacterial analogue (HMBA) to incorporate a portion of DNA from an arriving terrestrial bacterial genome, or for terrestrial bacteria (TB) to incorporate a portion of DNA from a hypothetical Martian bacterial analogue (Figure 1a).

Thus, from a genetic point of view, the donor does not have to be alive in order to be a donor. It does not even have to survive as an endospore. All that is required is that the recipient is alive, since natural competence for transformation is entirely controlled by the DNA-absorbing bacterium itself [Blokesch, 2016].

In terrestrial environments, destroyed bacteria can supply the exogenous DNA pool in the environment, where free DNA is often abundant. Exogenous DNA can in fact persist in the natural habitat for long periods of time and be viewed as representing a record of genes [Moradigaravand and Engelstädter, 2014]. Through DNA diffusion, these pools of DNA can potentially spread genes through a bacterial population, enabling a flow of genes without the need for any bacterial migration [Moradigaravand and Engelstädter, 2014].

It seems that most competent bacteria can take up available exogenous DNA from any source, although they take up DNA from their own species much better than DNA from distant bacterial species [Mell et al., 2014]. While the host cells bind double-stranded DNA at their cell surface, they nevertheless transport only single exogenous DNA strands into their cytoplasm [Chen and Dubnau, 2005]. However, it has been demonstrated that competent *Bacillus subtilis* are capable of incorporating the whole-genome DNA, that is, 4215 kb double-stranded DNA, from the protoplast lysate of *B. Subtilis subtilis* [Akamatsu and Taguchi, 2001], although this transformation was reported to be different from that incorporated by purified DNA conventionally [Saito and Akamatsu, 2006].

Thus, although the meteorite cannot protect its entrapped organisms at all 3 stages, it may be sufficient to protect the genetic material of the organisms until it arrives on the new world. This is quite relevant. While it is not yet known whether organisms can survive the journey between two worlds, it is almost a corollary of the former that such meteorites can carry fragments of life all the way to another world. Indeed, the probability that segments of RNA and DNA endure the journey seems higher than the probability of surviving bacteria.

Of course, RNA, DNA and other biotic components can be degraded before arriving at the destination. However, although such free genetic material can often be highly fragmented, terrestrial competent bacteria have evolved active DNA scavenging approaches that allow them to take up exogenous bases and nucleotides where possible [Blokesch, 2016].

This situation is thought-provoking. If it is assumed that a bacterium is indeed a universal blueprint for life, then it can be imagined that gene fragments from life elsewhere arrive on Earth. Since bacterial life exists in virtually all available niches on Earth, then bacteria will come into contact with this exogenous material. Thus, the HMBAs and TB could have their cellular material transferred to each other. Such genetic fragments can be of great importance, as they still represent complex information-carrying molecules. Thus, their incorporation can influence the native biological evolution.

Of course, this does not mean that such genes remain permanently in a gene pool. Certain genes can be driven to extinction. But, on the other hand, it can hypothetically provide an advantage, as the recombination of incoming DNA with the native DNA can affect the fitness of bacteria and become fixed in the gene pool.

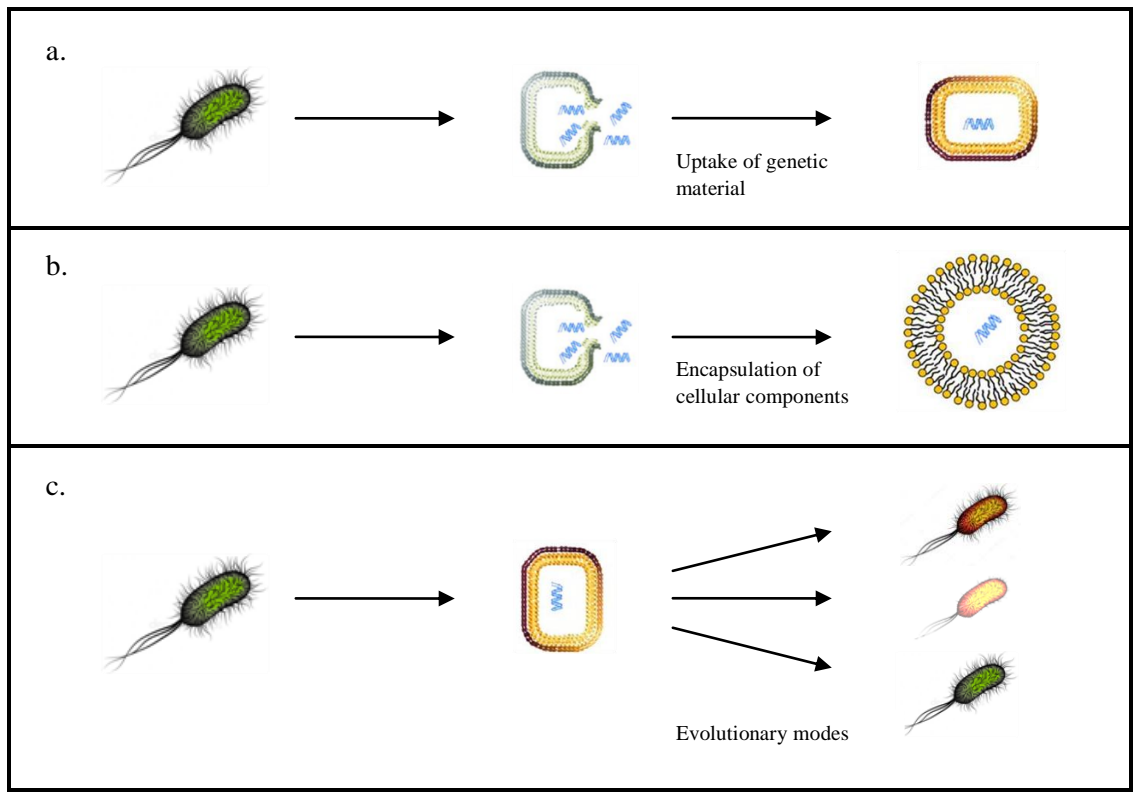

**Figure 1.** The 3 arrival modes of lithopanspermia (apart from no effect at all). (a) fractional genetic transfer, (b) fractional componential drift and (c) phenotypic continuance. Credits: adapted from Bravo et al, 2018; Zeineldin; and Mirumur, 2011, 2014.

**2.2. Fractional componential drift**

As discussed in the preceding section, bacteria do not *ipso facto* have to survive the transmission to another solar system body in order to have astrobiological significance. Thus, even if there is no native life where they are deposited by a meteorite, the bacterial components may still be important, if it is assumed that the meteorite lands on an acceptor world that has conditions suitable for life (Figure 1b).

The bacteria may have been viable all through the stellar transit, only to be destroyed upon arrival. Nonetheless, their machinery, including genetic systems, metabolic networks and numerous membrane lipids could be distributed in the environment surrounding the impact site. From a prebiotic point of view, the introduction of fragmented or intact machinery could be very important, as they still represent some of the most complex biostructures in existence. Such a contribution may help boost the native chemical evolution.

When lipid-like molecules are when placed in water, they can spontaneously aggregate to form lipid droplets, micelles, vesicles and bilayers even under simple conditions (Safran, 1994). The lipid bilayer is the basic structure for the terrestrial cell membrane. The structure and amphipathic nature of lipids allows entropy-driven hydrophobic interactions that enable the spontaneous formation of sealed compartments in aqueous conditions, and the resealing of tears [Alberts, 2002]. This spontaneous self-organisation into bipolar membrane sheets does not require complex chemical evolution. Instead, it depends entirely on physical characteristics of the molecules themselves [Segre et al., 2001]. Importantly here, the most abundant membrane lipid molecules are phospholipids, and that the bacterial membrane is often formed by one main type of phospholipid [Alberts, 2002].

Thus, even if bacteria are destroyed and their constituents distributed in the environment, their components may still become encapsulated and protected in the new environment. Such amphiphilic molecules presumably were abundant in the prebiotic environment on Earth [Segre et al., 2001]. These molecules are thought to have originated either from extra-terrestrial impacts or through abiotic synthesis [Pohorille and Deamer, 2009]. They may have also been abundant in the historically more aqueous environment of Mars. Amphiphilic molecules were found in the Murchison meteorite, which spontaneously formed lipid membrane vesicles when placed in water [Deamer and Pashley, 1989].

This scenario is initially merely a passive drift. Nonetheless, the components will be protected for a time in this prebiotic environment, which automatically takes advantage of what is randomly available. The origin of the first fully autonomous cell was probably not a singular event, but rather a chemical evolution that occurred through several stages. Research on the necessary aspects of a cell has resulted in 3 models for the origin of cells: the RNA-first model [Gilbert, 1986], metabolism-first model [Oparin, 1957] and lipid-first model [Segre et al., 2001]. Which of these truly operated has been a long-running debate. I will not presently discuss the pros and cons of these models. The salient point is that no matter how it proceeded, self-replicating genetic systems and metabolic networks must necessarily have been encapsulated within a lipid membrane compartment before the first true autonomous cell emerged. Thus, regardless of the order of development, encapsulation is an inevitable and crucial step in chemical evolution.

The arrival of exobiotic material and subsequent encapsulation on a new world does not fall directly into these 3 models. Rather, it is a form of hybridization between the lipids that are native to the acceptor world and the arriving exobiotic material from the donator world. Some of these lipids may have been constituents of the bacterial membranes. Phospholipids last longer than DNA and RNA, although they are degraded over geological time scales. This scenario thus does not ultimately solve the question of the origin of life. It simply shows what interplanetary transmission can do in principle.

If chemical evolution takes place on any world with the right conditions, such as the presence of liquid solvent; the so-called SPONCH elements (sulphur, phosphorus, oxygen, nitrogen, carbon and hydrogen) and energy sources, then the question will be what effect this can have. This situation would not be expected to occur on Mars, except in hypothetical sub-environments or in its Noachian era, where water seemed to have been abundant. But, for arriving bacterial analogues, it could potentially occur on Earth or any world with the right conditions. Whether such membrane-enclosed components will contribute to chemical evolution, accelerate it, initiate it even, or simply continue as fractional componental drift in the environment until they degrade remains at present speculative. For this discussion, it is only important to make clear that this possibility exists.

**2.3. Phenotypic continuance**

The scenarios discussed in sections 2.1 and 2.3—where it is the genotype and cellular components, or more precisely their fragments, which are important—are intriguing. But, it is perhaps even more interesting to discuss what happens if the organisms arrive alive, in what could be considered a successful transmission. This is, in a manner of speaking, the phenotype that counts. Once the influence of planetary science is complete, what requirements does evolutionary biology then set for the survival of life? What predictions can an evolutionary framework provide?

It can be argued that the arrival of organisms from one world to another through the stages of lithopanspermia can in principle be considered invasive biology. Terrestrial invasive biology is traditionally viewed as the invasion of a foreign species into habitats that are already inhabited. This species introduction effort was defined as 'a composite measure of the number of individuals released into a region to which they are not native' [Lockwood et al., 2005].

In the present context, interplanetary invasive biology can be considered as a process where a species from one world seeks to gain a foothold in a new world habitat, irrespective of whether or not life exists there already. Invasion of species into other environments or expansion of their ecological range into new environments is a natural phenomenon. An equilibrium balance between habitats is always a short-term state. Sooner or later the equilibrium changes and life spreads into other habitats. Thus, the arrival of life from one world to the next and the subsequent attempts to gain a foothold there is in principle no different.

On a planet, when expanding into new habitats, invading organisms rarely are successful in colonisation on the first attempt. In many cases, these initial invaders will not survive in the new environment. Usually several attempts are needed in a process termed propagule pressure before they gains a foothold.

Propagule pressure is a single consistent correlate of establishment success, and it clarifies that smaller populations face a greater likelihood of extinction than larger populations [Akcakaya et al., 1999]. Hence, consistent release of individuals, or the release of large numbers of individuals, into a new environment is necessary for survival and colonisation of the new environment [Lockwood et al., 2005]. The fact that the probability of success is low however does not mean that it cannot be done. It is still a non-zero possibility.

There is a crucial difference between this terrestrial situation and the situation brought about by lithopanspermia. While for invasive species on a planet like Earth, one member of the species after another enters the new habitat until they gain a foothold through propagule pressure, a species in the practical sense only arrives once onto another solar system body.

The arrival of life into a new world thus represents low propagule pressure. Even if meteorites arrive with a certain frequency, the frequency between arrivals will be millions of years. Even assuming that meteorites arrived within a few years of one another, they will still arrive at widely different locations on the acceptor world. So, in an evolutionary context, it will have the same effect as a single arrival, a propagule singulus.

This is important to mention because of the direct relationship between success and pressure. The higher the propagule pressure, the higher the possibility of success of the invasion. Therefore, a species from a donator world has to survive and gain a foothold in the acceptor world environment in the first attempt, since it will not be the case that it gradually moves into the new environment in the form of propagule pressure. Instead, a species will be in the same environment and exposed to that environment all at once.

This all-or-nothing response is not specifically due to the arrival of the meteorite. On worlds with atmospheres, entry proceeds at terminal velocity, that is, the constant maximum velocity reached by an object falling through the atmosphere under the attraction of gravity, which is approximately 50 m/sec for Earth and 300 m/sec for Mars [Nicholson, 2009]. Thus, the forces generated during the delivery of life on an acceptor world like Mars are in fact relatively modest compared to those prevailing when life was collected on a donator world, such as Earth. The point of this discussion is the severity of the environmental stressors the organisms will encounter in the environment on the new solar system body they arrive at. If these stressors are severe enough, then the arriving life could perish at the time of contact.

Of course, there may be millions of bacteria entrapped and protected in the meteorite that can have survived the arrival. But the situation will still be different from that of invading organisms on Earth. While terrestrial invasive biology so to speak has a 'back-up' supply of organisms to repeatedly try to gain a foothold, this does not apply to interplanetary invasive biology where the millions of organisms still count as a single attempt to gain a foothold.

**2.3.1. Evolutionary modes of response**

As discussed in section 2.3, an all-or-nothing response will apply when life arrives at a new solar system body. To make the situation even more complex, this interplay between planetary science and evolutionary biology shows that there exists an inverse proportionality between the exchanges of life between the two worlds. This is based on the fact that there are many different types of habitable environments in a world from which a sample of life can be obtained. However, there is only one type of habitable environment in another world to which this sample of life can be deposited.

Thus, the following rule can be stated: The retrieval of a particular sample of life in a random type of habitable environment on a donator world is inversely proportional to the deposition of a random sample of life in a particular type of habitable environment on an acceptor world. Therefore, the 'impactor-transporter-bringer' mechanism that successfully retrieves, transmits and deposits life is an invariant.

This rule states that organisms are adapted to the specific environment in which they live. The asteroidal or cometary impactor can land anywhere on a donator world and eject organisms. However, as these organisms had adapted to a specific environment on the donator world, it is not irrelevant where on an acceptor world the meteorite deposit them upon landing. The meteorite has to deposit them in a similar environment in order for them to survive. Thus, the impactor-transporter-bringer is considered a constant $k$ as long as it can successfully retrieve, transmit, and deposit organisms on another world.

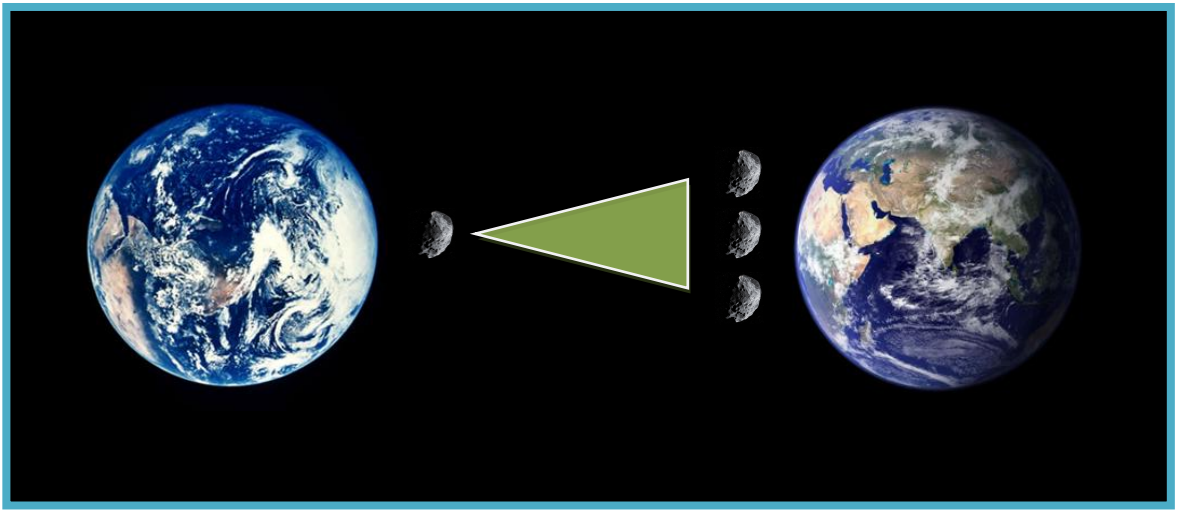

**Figure 2.** Rock transmissions between two Earth-like planets in a solar system given by an arrow. The tip of the arrow shows impact and ejection (the rock is not to scale) in an Antarctic-like environment. The base of the arrow indicates several possible impact sites for the same rock, in this illustration an ocean, a desert and a forested region. Credits: Earth pictures adapted from NASA.

From a purely physical viewpoint, the situation is the same. An impactor hits a single point on a solar system body and sends material upward. An impactor hits a single point on another solar system body and deposits material. This can be drawn as a line with a single point at each end. Thus, it could be inferred that if multiple environments can potentially support life, then it should be irrelevant from which of them life is retrieved from and deposited on.

However, planetary ejection and planetary entry are only 2 factors in the equation. Another factor is the evolutionary strategy of the transported organisms. Thus, the situation is not the same for the evolutionary strategy of the transported organisms. Earth and Mars serve as an example. Earth is a life-friendly planet, while present-day Mars overall is not, though it is may have sub-environmental niches where life could live. Thus, hypothetical Martian bacterial (or archaeal) analogues transported to and deposited on Earth could, from a planetary science perspective, be more likely to survive than the opposite scenario. Indeed, we could simplify the situation and look at a transmission of life between two planets with Earth-like conditions, as seen in Figure 2, which should presumably make it easier for life to gain a foothold.

However, from an evolutionary perspective, this is not valid. While it is correct that terrestrial life would have to be placed in sub-environments on Mars in order to survive, this does not guarantee success. Organisms are adapted to their environment. This includes even extremophiles. Thus, from a biological point of view, the situation is different from the physical one, given more by an arrow than a line, as in Figure 2. For example, two environments on Earth, the Sahara and Antarctica, are harsh, yet both contain several life forms. However, if organisms from a Sahara-like environment were deposited in an Antarctic-like environment, then their chances of survival will be low, even though both environments are capable of supporting life. Although one might be tempted to assume that extremophiles will be better suited to colonise other worlds, the above example indicates that this is not be valid.

Thus, there will be variation between two environments, even if they from an initial estimate appear to be similar. Even HMBAs, which upon arrival on Earth are more likely to be deposited in a place where there is already life than in the reverse scenario, will not necessarily be adapted to the environment.

It is possible that bacteria can only survive the journey in the form of, for example, desiccation tolerant life or bacterial endospores. The latter has long been considered the most durable form of known terrestrial life, capable of withstanding severe environmental stressors in the dormant state [Horneck et al., 2008]. For endospores, the difference between the environments will not be as influential. Endospores are essentially a dormant form encased in a nearly indestructible shell, but they do not do mush in themselves. To have biological significance, as a viable organism with a life cycle featuring growth, metabolism and reproduction, they must necessarily become active bacteria again on the acceptor world. This requires them to face the environment.

Thus, with both an inverse proportionality and an all-or-nothing response between changing environments, it seems that the possibility of a successful lithopanspermia, where a bacterial colony arrives in a potentially habitable environment on a new solar system body and automatically survives and adapts to it, is extremely challenging. However, terrestrial populations often encounter unpredictable and variable environmental conditions, and surviving under such fluctuating conditions poses evolutionary challenges [Martín et al., 2019]. A number of evolutionary strategies to cope with such environmental uncertainty have been proposed. Overall, there are 3 evolutionary modes of response, besides extinction, to fluctuating terrestrial environments: adaptive tracking, adaptive phenotypic plasticity and bet hedging [Simons, 2011] (Figure 1c). These 3 modes do not give the same probability of survival for the same environment. So, the question is how these evolutionary modes will come into play in an interplanetary transmission between two different environments. This will be addressed in the following.

**2.3.1.1 Adaptive tracking**

Organisms in changing environments experience selective pressures and can adapt to the new environmental conditions through adaptive tracking [Rago et al., 2019]. Thus, populations that experience environmental change will gradually adapt one or more traits. This means that the optimal trait values change continuously and that natural selection results in the gradual evolution of more fitting phenotypes, thus removing suboptimal forms that hitherto were well adapted to the environment [Simons, 2011]. This evolutionary mode of response requires that the necessary genetic variance exists, as well as mutational variance for fitness with respect to changing environments [Simons, 2011]. This is perhaps the most well-known mode. Any organism is adapted to its environment. When organisms spread to new environments, they become adapted as they spread, and over time can become a new species.

However, the transmission of TB to a sub-environment on Mars or HMBAs to Earth will not benefit from this mode, as it will not prepare the organisms for the journey itself, or for the initial encounter with the new environment. In this situation, the meeting between two environments is abrupt and extreme. There is no possibility of propagule pressure or time for the organisms to gradually adapt to the new environment. They must survive in the first attempt.

The main issue here is that environmental variance may lead to variable selection, meaning that traits that are optimal at one time or place are disadvantageous at another time or place. Organisms have adapted to a specific environment among many in the donator world, but here they arrive to a random environment on the acceptor world. This is the basis of the inverse proportionality discussed in section 2.3. However, the probability of survival can be high if the organisms happen to be adapted to the environment they arrive at on the acceptor world.

But what is the probability of that occurring? To model this situation, the binomial distribution formula can be used, where the probability of obtaining exactly k successes in n trials can be calculated. Thus, the following simplified situation can be set up. A bacterial colony encounters 10 different environments in an acceptor world. If each environment is either similar or non-similar to the colony's native environment, what is the probability of encountering exactly 1 environment similar to the native environment on the donator world? Plugging in the numerical values:

n = 10
k = 1
n - k = 9 = number of failures
n - k = p = 0.5 = probability of encountering a similar environment
1 - p = q = 0.5 = probability of encountering a non-similar environment,

the probability that the colony encounters 1 environment, and also does not encounter a specific set of 9 environments, is calculated as:

$$P_{\text{specific 1 environment}} = (0.5)^1 \cdot (0.5)^9 \qquad (1)$$

However, it is necessary to update this based on how many ways it is possible to divide a group into sets of 1 and 9, which is obtained by the binomial coefficient. Thus, the final probability is:

$$P(\text{match out of 10 environments}) = \binom{10}{1} \cdot (0.5)^1 \cdot (0.5)^9 \tag{2}$$

$$\approx 0.0098 \approx 1\%.$$

Thus, it is possible for the bacteria to survive the encounter with a new solar system body through the evolutionary mode. However, the probability is not encouraging, especially since only 10 environments are a generous assumption (Remember, we have restricted ourselves only to environments where the possibility of life is present, and such worlds will in principle have many different environments). The probability that the organisms are not adapted to the specific environment they arrive in is much higher than the opposite case.

If a random environment allows any possibility of life, then it would be possible to adapt to it. But adaptive tracking requires a period of time to take place. Upon the arrival of organisms from a donator world to an acceptor world, their colonisation of the environment will not take place during a long period of time with the possibility of adaptation and multiple attempts. Rather, it will be virtually instantaneous. Even terrestrial extremophiles adapted to a cold, arid desert environment can face serious challenges if transported to a sub-environment on Mars. As soon as there is an opening through to the meteorite where the organisms have lived protected, the organisms will be in contact with the new environment and the reaction will be virtually instantaneous. Their survival will be an all-or-nothing response.

Upon arrival, the meteorite will likely be further shattered, and it and its cargo will effectively be distributed into the acceptor worlds crust, water or ice. But even though the incoming bacteria are distributed over a larger area, collectively it is still an all-or-nothing response that applies, since the arriving organisms as a clonal colony are similar.

As stated, this mode can, over time, allow the organisms to adapt to this new environment. But the issue here is that they do not have that time, because adaptive tracking requires time to adapt to a new environment. There is an unavoidable time lag in adaptation behind shifts in optimal trait values [Simons, 2011]. If the rate of environmental alteration is sufficiently slow, as might be the case in the native environment, the mean phenotype will remain close to the optimum such that a fitness depression can be avoided [Lynch and Lande, 1993]. But if the environmental alteration is too sudden, or the new environment is alien, as is the case here, the organisms will be overwhelmed. Even if the organisms survived the initial encounter, then due to this time lag in adaptation, they will become extinct since there is no time for such adaptation to the environment, even for organisms like bacteria that can mutate rapidly with high natural selection rates.

Of course, these bacteria may have survived the journey in the form of endospores or microbial cysts. This type of protection can also protect them when the meteorite is shattered upon impact. But the fact remains that if the bacteria are to have a life cycle on the acceptor world, then they have to emerge from their protection and face the new environment.

This evolutionary mode, in this situation with a shift to an alien environment, seems most suitable before planetary launch, since it may have evolved organisms that can survive the journey. It can also appear to be suitable for colonising the new environment after surviving the encounter with the new environment.

However, there is one further limitation of this evolutionary strategy under fluctuating selection. Heritability itself is environment-dependent, and negative correlations exist between the two parameters of heritability and the strength of natural selection [Wilson et al., 2006]. Responses to natural selection will in fact be constrained and phenotypic stasis will be favoured when environmental shifts are too harsh. Thus, environmental variance reduces heritability under strongly unfavourable conditions, which certainly is the case here with the shift to an acceptor world with its potentially different environmental conditions.

The collective considerations indicate that the evolutionary mode provides little opportunity for bacteria to gain a foothold on another solar system body.

**2.3.1.2 Adaptive phenotypic plasticity**

This evolutionary mode makes it possible for individuals to achieve an adaptive fit between phenotype and environment through the modulation of phenotype in response to direct environmental sensing [Beaumont et al., 2009]. It is an effective solution to environmental variance, since it instantly attains the most fitting phenotype for a range of conceivable environments [Simons, 2011]. Adaptive phenotypic plasticity is thus superior to the previous mode for a wide range of environmental conditions [Rago et al., 2019].

The initial situation is the same as in the previous section. As soon as there is an opening through to the meteorite, the organisms will be in abrupt and extreme contact with the new environment. There is no possibility of propagule pressure in the new environment. They must survive in the first attempt.

This mode appears to give organisms from a donator world a higher probability than the previous mode in surviving the initial encounter with the new environment on an acceptor world since it is not so limited by the either-or-situation with traits that are advantageous at one time but disadvantageous at another, but instead has a range of phenotypes expressed over a range of environments. However, the evolution of adaptive phenotypic plasticity does have a serious limitation in that it requires the phenotype-fitness association to be predictable across the native environments. Organisms utilising this evolutionary strategy produce only optimum fitness phenotypes by responding appropriately to environmental cues [Reed et al., 2010]. Thus, it is an effective strategy only under the range of native environments previously encountered by the arriving bacteria.

Obviously, as a remark only, since impacts can occur at such large intervals of time, none of which follow a predictable sequence, and can occur at any point on a world, they are, from an evolutionary point of view, an unpredictable event (though from a physical point of view they are not). There are no cues available for this event.

Adaptive phenotypic plasticity is adaptive only if there are reliable cues that allow the arriving organisms to match the phenotypes to the environment of the acceptor world. The arriving organisms will only be able to easily survive the initial encounter with the new world if they have experienced cues about a similar environment on the donator world. The issue here is that the arriving organisms have evolved to diverse, but still restricted predictable environments on the donator world, and here they arrive to an environment on an acceptor world that can be any random environment.

Furthermore, it has been shown that when the correlation between cue and optimum is weakened and environmental variability is high, strong plasticity diminishes the population size, so that it faces a markedly higher probability of extinction [Reed et al., 2010]. Both the journey and the encounter with the new world are obviously an extreme situation. Even if individual organisms survived, their continued survival through this mode itself will be further reduced.

Thus, there is an inverse proportionality here, as was discussed in section 2.3. Although the bacteria possess a range of phenotypes expressed over a range of environments, phenotypic plasticity is pragmatically in the same situation as adaptive tracking. Thus, their probability of arriving at an appropriate environment and surviving is again given by equation 2, with P = 0.0098 or 1%. As for adaptive tracking, adaptive phenotypic plasticity is constrained. This evolutionary mode in this situation with the shift to an alien environment would be most suitable before planetary launch.

Taking this information together, the evolutionary mode has greater range, but nevertheless provides little opportunity to gain a foothold on another solar system body.

**2.3.1.3 Bet hedging**

In this evolutionary mode of response, bet hedging traits can evolve under conditions of unpredictable environmental variance, where long-run or expected geometric mean fitness is maximized over time, even at the cost of a decrease in the arithmetic mean [Simon, 2011]. Fitness is treated here as a random variable, where the fitness of individual organisms is not known in advance, but where the fitness can nevertheless be treated by a probability distribution [Starrfelt and Kokko, 2012]. Thus, this evolutionary strategy can be viewed as an adaptation to unpredictability itself [Simons, 2011].

The initial situation is the same as in the previous sections concerning the abrupt and extreme contact with the new environment upon meteorite impact and an all-or-nothing response.

This arrival at another solar system body can clearly be viewed as unpredictable natural selection. Bet hedging is an evolutionary strategy that generates stochastic variation in fitness-related traits among TB or HMBAs. This strategy in essence distributes the risk among an array of phenotypes, each one neither optimal nor a failure across multiple environments. This increases the probability that some of the arriving bacteria express a phenotype that will ensure their immediate survival in this new unpredictable world.

The inverse proportionality as discussed in section 2.3 still exists. But bet hedging is the evolutionary strategy that seems best to smooth out this inverse proportionality between any point on a donator world versus one point on an acceptor world. This is because this mode of response can ensure survival under the broad array of environmental situations, in contrast to the more restricted adaptive tracking and adaptive phenotypic plasticity.

The previous sections detailed how the adaptive trait or the environmental cue of the bacteria had to match the new environment to permit survival. Here the situation is different. Bet hedging does not require a one-to-one match. Rather, it has a wider range over environments. Thus, it should not be asked what the probability is regarding the effectiveness of this mode on exactly one of 10 environments on an acceptor world. Instead, the correct wording should be that if an evolutionary mode is effective in, for example, 75% of the time in shifting environments on a donator world, what is the probability that the mode will be effective on, for example, exactly 6 of 10 environments on an acceptor world? Plugging in the numerical values:

$n = 10$
$k = 6$
$n - k = 4$
$p = 0.75$ = probability of effectiveness
$q = 0.25$ = probability of ineffectiveness,

the probability that this mode works on a specific set of 6 environments and also does not work on a specific set of 4 environments is:

$$P_{\text{specific 6 environment}} = (0.75)^6 \cdot (0.25)^4 \tag{3}$$

However, as before, it is necessary to update this calculation based on how many ways it is possible to divide a group into sets of 4 and 6, which is obtained by the binomial coefficient. Thus, the final probability is:

$$P(\text{effectiveness on 10 environments}) = \binom{10}{6} \cdot (0.75)^6 \cdot (0.25)^4 \tag{4}$$

$$\approx 0.1459 \approx 14.6\,\%.$$

The probability is not high (given the specific assumptions), demonstrating that transfer to another solar system body and survival upon arrival are drastic processes. The 75% calculated effectiveness is based on the fact that the bacteria present comprise different phenotypes, with different possibilities of survival, in contrast to the first mode, where the same number of bacteria were similar. The 4 environments may have environmental stressors that are simply too far from what bet hedging evolved to deal with, or are too severe for all life. However, the interrelationships between the probabilities of the different modes remain relatively similar, with bet hedging yielding the highest probability.

The organisms the meteorite brings to a new world did not develop their bet hedging strategy with this interplanetary transit as the reason. Instead, the meteorite brings with it organisms that have obtained bet hedging on the donator world. To illustrate how bet hedging strategies can evolve on a donator world, I will use a simple model that follows a methodology originally advanced by Seger and Brockmann (1987), and then discuss its application when meeting an acceptor world. A bet hedging strategy can evolve by assuming that, for example, the native environment of the donator world can be either warm or cold during a generation, where bacteria can encounter warm and cold periods of time with equal frequency.

It holds that none of the bacterial genotypes can change their development depending on the native environment they live in and since it is a bacterial colony asexual inheritance is in effect. The reproductive success of each bacterial genotype in each period listed in Table 1 is equivalent to the absolute fitness of an individual bacterium [Seger and Brockmann, 1987; Starrfelt and Kokko, 2012].

**Table 1.** Absolute fitness values for 4 different genotypes.

|  | $A_{cold}$ | $A_{warm}$ | $A_{con}$ | $A_{div}$ |
|---|---|---|---|---|
| Cold, $P_{cold} = \frac{1}{2}$ | 1 | 0.560 | 0.765 | 0.745 |
| Warm, $P_{warm} = \frac{1}{2}$ | 0.540 | 1 | 0.765 | 0.815 |
| Arithmetic mean fitness $\delta$ | 0.770 | 0.780 | 0.765 | 0.776 |
| Geometric mean fitness $\Omega$ | 0.735 | 0.748 | 0.765 | 0.775 |

These values reflect several assumptions. First, a constant of 100 means that a bacteria of genotype $A_{cold}$ will produce 100 descendants of the same genotype in a cold period (in reality, it would be 128 due to the constancy of bacterial doubling, which for simplicity is assumed here to be 100), which is equivalent to 1, and 54 descendants in a warm period of equal duration as the cold period, which is equivalent to 0.540. Secondly, genotype $A_{warm}$ bacteria will produce 100 descendants of the same genotype in a warm period, which is equivalent to 1, and 56 descendants in a cold period of equal duration, which is equivalent to 0.560. In this model, the fitness of, for example, $A_{warm}$ bacteria is the fitness achieved in cold and warm periods, where the probability of a period being cold is P = 1/2 [Seger and Brockmann, 1987; Starrfelt and Kokko, 2012].

Now consider the situation in the native environment where the population was initially fixed for $A_{cold}$, in which an $A_{warm}$ mutant has just emerged. Thus, comparing the two specialists, $A_{cold}$ and $A_{warm}$, their arithmetic mean fitness is calculated as:

$$\delta_{cold} = P(\text{cold period}) \times (0.540) + P(\text{warm period}) \times (1) = 0.770, \qquad (5)$$

and

$$\delta_{warm} = P(\text{warm period}) \times (0.560) + P(\text{cold period}) \times (1) = 0.780. \qquad (6)$$

The geometric mean fitness, for the two specialists is calculated as:

$$\Omega_{cold} = (0.540 \times 1)^{P(\text{cold period year})} = 0.735, \qquad (7)$$

and

$$\Omega_{warm} = (1 \times 0.560)^{P(\text{warm period})} = 0.748. \qquad (8)$$

In this methodology there is temporal variability in the native environment, whereas there is no spatial variability [Seger and Brockmann, 1987]. This means that all bacterial individuals experience either a warm or a cold period, which in turn means that since $A_{warm}$ has the highest geometric mean fitness, it will eventually prevail, that is, increase to fixation.

Thus, if $A_{warm}$ arrives on an acceptor world in our lithopanspermia scenario, it could indeed survive if it encounters an environment that is either cold or warm (it is assumed that the temperatures are within the limits for sustainable bacterial life), or an environment that can vary between cold and warm, and thus match its native environment. In this scenario, it is a specialist whose situation is virtually similar to the situation in adaptive tracking. The same is true for $A_{cold}$, which is also a specialist whose situation is virtually similar to that of adaptive tracking.

In another scenario, on the donator world, $A_{warm}$ is fixed and $A_{con}$, the conservative bet hedger, emerges as yet another mutant. This is a consistent but low risk phenotype within a genotype [Liu et al., 2019]. Thus, $A_{con}$, is a generalist, that here is assumed to have a fitness of 0.765.

Bacteria have historically been viewed as clonal populations of identical prokaryotes, which result from symmetrical cell division. The cells are genetically identical with their phenotypes merely reflecting the genetic constitution [Casadesus and Low, 2013]. However, this view is not valid. Besides the fact, that *Bacillus subtilis* form spores [Errington, 2003] and *Caulobacter* undergo asymmetric cell division [Kirkpatrick and Viollier, 2012], which represent bacterial development where the genome sequence remains unchanged, there are also phenotypically distinct bacteria [Casadesus and Low, 2013]. Various mechanisms involved in bet hedging strategies in bacteria include phase variation [van der Woude et al., 2004], contingency loci [Moxon et al., 2006] and epigenetically inherited bistability [Veening et al., 2008].

$A_{con}$ specialists do well regardless of the native environmental conditions. They are equally unremarkable in both periods and are typically less fit than the $A_{cold}$ or $A_{warm}$ specialists on average. Because bet hedging adaptation sacrifices short-term performance, it appears detrimental compared to the other strategies. However, $A_{con}$ has higher geometric mean fitness than $A_{warm}$, because its fitness does not vary. $A_{con}$ bacteria eventually prevail and colonise [Seger and Brockmann, 1987; Starrfelt and Kokko, 2012]. Thus, although $A_{con}$ bacteria are characterised by lower arithmetic mean fitness, they will still replace a population of $A_{cold}$ bacteria and $A_{warm}$ bacteria, because the individual geometric fitness of the latter two types are lower than $A_{con}$. One period out of two, each specialist suffers the cost of being maladapted to the environment, while $A_{con}$ fitness does not vary and it will continue in every period.

Thus, in our lithopanspermia scenario, if the population has gone to fixation in the native environment of the donator world, then only $A_{con}$ will arrive with the meteorite to the new environment on the acceptor world. Since its success results from the lack of pronounced success in any period and avoiding very low success in any period in its native environment, then it will, regardless of whether it encounters a cold or warm environment, be able to handle the initial brunt from the new environment. The conservative generalist $A_{con}$, with its reduced genotypic variance, will thus do reasonably well regardless of the environmental conditions of the acceptor world.

If the arriving population has not yet gone to fixation, then $A_{cold}$, $A_{warm}$ and $A_{con}$ individuals will be able to arrive to the new environment. They will experience the full brunt of the conditions of the acceptor world's environment. $A_{cold}$ and $A_{warm}$ specialists will each suffer the cost of being maladapted and will essentially be in the situation of adaptive tracking. In contrast, $A_{con}$ fitness does not vary and so these bacteria will persist.

If $A_{cold}$ and $A_{warm}$ individuals survived the initial encounter, they will still be able to survive for a while. Yet neither would continue to survive against $A_{con}$ because the geometric fitness of each individual is lower. Eventually $A_{con}$ will outcompete $A_{cold}$ and $A_{warm}$. Thus, $A_{con}$ is the biological equivalent of a 'jack-of-all-trades and master of none'. This is precisely the quality that allows it to displace a mixture of specialists, such as $A_{cold}$ and $A_{warm}$, under temporal environmental variation. Furthermore, it is precisely this quality that gives the bacteria an advantage over the specialists upon arrival in a new world, regardless of whether the environment is warm or cold. Finally, with $A_{con}$ fixed in the native environment, we can consider the emergence of yet another mutant, the diversified bet hedger $A_{div}$. The diversified bet hedging strategy reduces the correlation of reproductive success between bacterial individuals sharing the same allele, which reduces the genotypic variance. Thus, $A_{div}$ individuals picked at random will not all display the same reproductive success within a period. Rather, some individuals will have developed into $A_{cold}$ bacteria, while other individuals will have developed into $A_{warm}$ bacteria [Seger and Brockmann, 1987; Starrfelt and Kokko, 2012].

The diversified bet hedging strategy's fitness values are calculated by assuming that the cold period specialist is produced with a probability of 0.42. Thus, the $A_{div}$ genotype develops the phenotype of cold period specialists 42% of the time and the phenotype of warm period specialists 58% of the time. Thus, the values are calculated as:

$$\text{Fitness}_{cold} = (0.42 \times 1 + 0.58 \times 0.560) = 0.7448, \tag{9}$$

and

$$\text{Fitness}_{warm} = (0.42 \times 0.540 + 0.58 \times 1) = 0.815. \tag{10}$$

The arithmetic mean fitness of $A_{div}$ in the native environment is given by:

$$\delta_{div} = P(\text{cold period}) \times (0.42 \times 1 + 0.58 \times 0.560) + P(\text{warm period}) \times (0.42 \times 0.540 + 0.58 \times 1) \tag{11}$$

$$= 0.776,$$

and the geometric mean fitness of $A_{div}$ in the native environment is given by:

$$\Omega_{div} = (0.42 \times 1 + 0.58 \times 0.560)^{P(\text{cold period})} \times (0.42 \times 0.540 + 0.58 \times 1)^{P(\text{warm period})} \tag{12}$$

$$= 0.775.$$

Thus, the $A_{div}$ genotype obtains a higher geometric mean fitness than the 3 previous genotypes, and even drives $A_{con}$ to extinction. If we re-examine our lithopanspermia scenario, then it follows that if the arriving bacteria have not yet gone to fixation on the donator world, then both $A_{cold}$, $A_{warm}$, $A_{con}$ and $A_{div}$ individuals will be able to arrive on the acceptor world. They will all now experience the full brunt from the new environment. $A_{cold}$ and $A_{warm}$ will now each suffer the cost of being maladapted. $A_{con}$ fitness does not vary and it will go on, while $A_{div}$ will continue to produce both cold period and warm period specialists with a fixed probability.

The phenotype of $A_{div}$ offspring develops independently of both the progenitor cell and the new environment on the acceptor world. These individuals can thus be viewed as 'flipping' a coin to determine which phenotype to develop into [Starrfelt and Kokko, 2012].

$A_{cold}$ and $A_{warm}$ individuals will still be able to survive for a while, yet neither $A_{cold}$ nor $A_{warm}$ would continue surviving against $A_{con}$ and $A_{div}$, because the geometric fitness of each individual is lower. If this took place in the colony's native environment, then $A_{div}$ would eventually outcompete $A_{con}$ and go for fixation, because one or the other of its phenotypes will be highly successful, and there is in fact never a period where the genotype generally does poorly. Indeed, if the $A_{div}$ bacteria in the population had gone into fixation upon arrival on the new world, then both cold period and warm period specialists would continue to be produced, which would allow the population to cope with the initial brunt from the new environment, regardless of whether it was a cold or warm environment, since it can produce specialists for both a cold or warm environment at the same time.

It could be said that the acceptor world environment, unlike the donator world environment of the arriving organisms, does not experience a shift between cold and warm, thus representing a new and interesting situation for bet hedging. The advantage of the $A_{div}$ phenotype is that it evolves in an environment that changes periodically between cold and warm. However, if the environment on the new solar system body remains either cold or warm, then $A_{div}$ bacteria will suffer the total cost of continuously producing many individuals (e.g., warm-adapted individuals that are maladapted to the cold environment), while the fitness of $A_{con}$ bacteria does not vary and these bacteria will do reasonably well regardless of circumstances, and so will persist.

However, although $A_{con}$ is a genuine bet hedger, it has not evolved optimum traits and is thus competitively inferior. Thus, although $A_{div}$ produces both cold period and warm period specialists and suffers the cost of some of these being maladapted, specialists with optimal traits are still produced. Thus, it will still outcompete $A_{con}$, especially in a world without changing environments, in which the environment may eventually prove too harsh for $A_{con}$. By producing high-fitness individuals, $A_{div}$ will still be able to remain in the new environment.

As discussed, $A_{cold}$ bacteria could indeed survive if they arrive in an environment that matches their native environment. This is a case of a specialist whose situation closely matches the situation in adaptive tracking. However, the probability of this happening is low. If there are 10 different environments on the acceptor world, then the probability that bacteria encounter its own environment is P = 0.0098 or 1%.

However, it could be objected, that if $A_{cold}$ bacteria arrive in an environment suited for $A_{warm}$ bacteria, which is not the optimal environment, it will still be able to produce 54 descendants in a warm period, rather than the 100 descendants in a cold period. Thus, it will suffer the cost of being maladapted overall, but some individuals will be able to survive. While it may ultimately become extinct, it can survive the initial encounter and subsequent encounters for some time, thus defying the above probability. The same would be the case for $A_{warm}$.

However, the issue here is that the model discussed is too simple. The model features a probability of P = ½ to encounter the right environment, where either $A_{cold}$ or $A_{warm}$ could be optimally fit. This is a very generous assumption. In reality, there will be more than just two environmental stressors (a and b) that can change in a given environment. This can be seen during, for example, bacterial infection [Casadesus and Low, 2013], during which the immune system mounts more than just two environmental stressors against the invading pathogens. Despite the numerous responses of the immune system, it is well-known that certain bacteria can utilize phenotypic switching in order to survive in the body much longer than bacteria that do not use this strategy.

The foregoing means that a given bet hedger would have to deal with, for example, four environmental stressors (a, b, c and d), which could be interrelated. In this example, c and d could be specific nutrients, specific pH, wet, dry environments or other features. It could the case that a ≠ b, which in a applies to c > d, and which in b applies to c < d, where the mutual relationships between c and d in both a and b vary from year to year.

Here the situation is more complex. In their native environment, $A_{cold}$ and $A_{warm}$ have evolved fitness with respect to the exact relationship of a ≠ b. The probability that $A_{cold}$ or $A_{warm}$ encounter an environment on the acceptor world with all the parameters present in the right relations is likely lower than P = ½. It would be more difficult for these bacteria to gain a foothold on the new world.

However, even in this more complex situation, the situation would still proceed the same way regarding bet hedging. In the donator world environment, $A_{con}$ or $A_{div}$ would evolve fitness via bet hedging with regard to all four environmental stressors. Whether they have reached fixation or not when they arrive on the new world, $A_{con}$ or $A_{div}$ would be able to survive and thrive when they encountered an environment where all four environmental stressors (a, b, c and d) are in interplay. This is why bet hedging offers a greater possibility for survival of organisms transported to another solar system body.

An arriving meteorite may have harboured millions of bacteria that had evolved to behave as if they sensed a stressful environment. Even following the initial encounter with the environmental shock of the new world, a few bacteria may survive. The survivors might then proliferate and establish a new clonal colony of bacteria. This, of course, is precisely the reason why bet hedging is in place to begin with.

For this new growing population of bacteria, specific environmental cues may now be available. This creates the possibility for the evolved expression of an optimal bacterial phenotype over a range of new environments, rather than merely maintaining an array of bacterial phenotypes, with each phenotype being neither optimal nor a failure across new environments. Over time, adaptive tracking can become possible so that species formation will occur. The initially genotypically similar, but phenotypically diverse, population of bacteria will give rise to diverse organisms with optimal trait values in their new world. While propagule pressure did not matter when the bacteria arrived on the acceptor world, it may matter now. The population of bacteria will be located at the site where the meteorite deposited them, and as they or their evolved descendants begin to expand their range, population pressure will come into effect again.

There is another situation that needs to be addressed. The overall goal for arriving bacteria is to survive the encounter with the new environment. This environment may be devoid of other life or may already contain native life. The environmental stressors can thus originate from the environment itself or from competing species. Even so, this phenotypic strategy is still convenient for the success of population expansion, regarding both the environment itself and competing species. Thus, populations with increased variability in individual risk-taking are capable of colonising wider ranges of territories [Martín et al., 2019].

HMBAs arriving on Earth during and after the Archean Eon would certainly have encountered native life forms. TB exist in virtually all available niches on Earth. Thus, arriving bacterial analogues will both have to survive the encounter with the new environment, and, assuming this life is life as we know it, and thus share traits with them, compete with the indigenous species for the resources or create a viable ecosystem in a new world. For the latter, it is important to make clear that while a world can be inhabitable for life, this does not necessarily mean that there is life on that world [von Hegner, 2019].

It is possible that despite this mode the arriving organisms do not survive the initial encounter. The probability of success is calculated as P ≈ 0.1459 or 14.6%. The model parameters are very general and can vary. The 4 environments where success did not occur denote environmental stressors that are simply too different or severe for any type of life as we know it. Thus, there are still restrictions for this mode in an interplanetary context. Life has its limits. Still, bet hedging is still the evolutionary strategy that best smooths out this inverse proportionality between any point on a donator world versus one point on an acceptor world.

**3. Conclusion**

Astrobiology, as the name implies, requires a framework of both astronomy and biology. Thus, planetary science (and associated physical disciplines) is only one part of the equation. Evolutionary biology (and associated biological sciences) is the other part of the equation. While planetary ejection, interplanetary transit and planetary entry are essential to the transmission of life between solar system bodies, the quality, quantity and evolutionary strategy of the organisms are essential for organism survival and colonisation on a new world. A successful lithopanspermia requires a successful evolutionary strategy.

Lithopanspermia experiments demonstrate that not all organisms can survive the journey. Evolutionary theory also predicts that although organisms survive the journey, it does not mean that they can survive the encounter with the new world as well. Thus, evolution sets a hard boundary through the inverse proportionality between the two points on the donator and acceptor world, with lithopanspermia necessarily interacting with both point.

Although evolution sets the boundary through the adaptation of organisms to a specific environment, evolution has nevertheless evolved space for stochastic phenotypic switching, a loop hole so to speak. This provides an adaptive solution to life when facing uncertainty by generating diversity even among a colony of genetically identical organisms, allowing them to increase their robustness against environmental variation. This increases an organism's probability of survival when confronted by environmental shocks. Thus, phenotypic quality matters.

The bet hedging strategy has been suggested to be among the earliest evolutionary solutions to life in fluctuating environments on Earth [Beaumont et al. 2009]. It could represent the most fundamental strategy to ensure the survival of life on a new world. In fact, bet hedging may also contribute to survival during the transport between two solar system bodies. Thus, preadaptation is a common strategy in which a minority of the bacteria in a population utilizes stochastic phenotypic switching to adopt a dormant metabolic state. The consequent slower bacterial growth increases their robustness to environmental shocks [Ogura et al., 2017]. ´

The successful transmission of life can be understood in an evolutionary framework by realizing that although the transmission of life from one world to another is an extreme form of population expansion, it is fundamentally still a population expansion. If life can be transported to an environment in another world, then this belongs in an ecological setting. Although this is not a traditional setting, bet hedging theory, which has been applied in many other traditional ecological settings, will still be applicable to face this situation with an all-or-nothing response. Among the organisms involved, there must at least be one capable of survival immediately in this environment.

Given that meteorite transfers between worlds exist and that evolution is the guiding principle of life, it could be extrapolated that the transport and colonisation of neighbouring worlds is a naturally occurring astrobiological phenomenon in a solar system. This may be another means by which organisms secure their existence, by spreading the risk of existence over several worlds and exploring all available niches in a solar system.

Bet hedging may have already played an unrealised role in experiments relevant to astrobiology. Numerous experimental organisms have been subjected to simulated Mars-like conditions, as well as to simulated and true space conditions. The percentage of surviving organisms was measured in part by how long they were exposed to these conditions, according to the intensity of these conditions and other factors (for review, see e.g. de Vera et al., 2019). In experiments where the relationship between time and intensity were considered, the percentage of survivors should be relatively similar in repeated experiments among genetically identical organisms of the same species. Cases where a seemingly random survival over repeated similar experiments occurs might be attributed to statistical chance, with some organisms surviving purely due to chance. However, from a bet hedging standpoint, this latter case need not be attributed to statistical chance. This random survival of individual members may mean that the organisms involved have already evolved bet hedging strategies. Thus, a seemingly random survival is not actually so random. Future studies should address this.

Of course, this situation requires that bet hedging have evolved in the population transmitted and deposited by the meteorite. The asteroidal or cometary impactor may not necessarily transmit organisms that have this strategy. However, bet hedging is widespread among bacteria, and bacteria are the dominant life form on Earth, existing in virtually all niches. However, $A_{con}$ and $A_{div}$ bacteria may not have evolved to the parameters they encounter in the new world. Even with bet hedging, the new environment can be quite dissimilar from the environment where a bet hedging strategy evolved. Thus, $A_{con}$ and $A_{div}$ do not represent universal types that can live everywhere. They can survive a wider range of environments than specialists and the probability of their survival is thus higher.

However, these bacteria evolved in a particular environment where a number of parameters existed, and which must exist if they are to survive in a new environment. So the inverse proportionality still exists. These facts indicate that although organisms invading and colonising neighbouring worlds may be a naturally occurring astrobiological phenomenon in a solar system, the probability of successful transmission of life to another world is low. These considerations will also have implications regarding the accidental propagation of terrestrial life to other worlds brought about by space exploration, and the discussion of planetary protection.

I have treated this as a one-way scenario. But it could also be treated as a two-way scenario. Thus, life may have been transported to another world, and descendants of that life may since have been transported back to the first world. Since life evolved in the different environments in the intervening time, such life will not necessarily be the same.

If life has indeed been transported back and forth over multiple rounds between, for example, Earth and Mars, then from a biological perspective these planets are considered two ecosystems interacting with each other. They are a simplistic dual biosphere. Mars may have been habitable during its Noachian era [Worth et al., 2013], and life may have existed there in the past. Indeed, it may even exist in the present in the form of an analogue to the terrestrial organism *Halorubrum lacusprofundi*. This is a microorganism that is both a halophile and a psychrophilic organism. In principle, the organism could be capable of living in Martian brines where it can survive under Martian surface temperatures [Reid et al., 2006].

As discussed, we could simplify the discussion and easily imagine a solar system where there are two or more Earth-like planets. Many solar systems exist in our galaxy, and the idea that a solar system with two worlds suitable for life exists is not far-fetched. Thus, here the effect of such reciprocal genetic exchange of altered gene pools could be profound. Such worlds should not be considered as isolated worlds, but instead as ecological niches where life can be exchanged. It is true that in such a general ecosystem it will not be easy to say whether life occurred on either or only one of them to begin with.

Indeed, if components produced by chemical evolution can move back and forth between two worlds, as discussed in the section on fractional componentional drift, chemical evolution could occur happening on two worlds. From an evolutionary point of view, it is better to perceive the origin of life as occurring in a dual world system rather than a one world system.

In summary, I have introduced an analysis to understand conditions happening in a colony that is expanding into a new planetary environment. The analysis provides a bridge between the theory of bet hedging, invasive range expansion and planetary science. Of course, as with all scientific models, bet hedging rests on a number of assumptions. Thus, the concept of geometric mean fitness relies on assumptions, such as nonoverlapping generations and externally fixed population sizes [Liu et al., 2019], which are open to research and possible modification. However, the results of the analysis highlight the demands and restrictions between population diversity and the environmental variations encountered during range expansion into a new planetary environment.

More work is needed. Applying evolutionary theory to the interplanetary exchange of life is a novel attempt. The generality of this framework makes it a valuable starting point for analyses and modelling of ecological scenarios regarding transmission of life or biocomponents between worlds in a solar system.